\documentclass[preprint,aps,showpacs]{revtex4}

\usepackage{graphicx}
\usepackage[all]{xy}
\newcommand{\be}{\begin{equation}}
\newcommand{\ee}{\end{equation}}
\newcommand{\ben}{\begin{eqnarray}}
\newcommand{\een}{\end{eqnarray}}
\newcommand{\wt}{\widetilde}
\begin{document}

\title{New soluble nonlinear models for scalar fields}
\author{C A Almeida,$^{a,}$\footnote{On leave from Departamento
de Matem\'atica, Universidade Regional do Cariri, 63.100-000,
Crato, CE, Brazil} D Bazeia,$^a$ L Losano,$^a$ and J M C
Malbouisson$^b$} \affiliation{$^a${\small Departamento de
F\'{\i}sica, Universidade Federal da Para\'\i ba, 58051-970,
Jo\~ao Pessoa, PB, Brazil}\\
$^b${\small Instituto de F\'\i sica, Universidade Federal da
Bahia, 40210-340, Salvador, BA, Brazil}}

\begin{abstract}
We extend a deformation prescription recently introduced and
present some new soluble nonlinear problems for kinks and lumps.
In particular, we show how to generate models which present the
basic ingredients needed to give rise to ``dimension bubbles,"
having different macroscopic space dimensions on the interior and
the exterior of the bubble surface. Also, we show how to deform a
model possessing lumplike solutions, relevant to the discussion of
tachyonic excitations, to get a new one having topological
solutions.
\end{abstract}


\maketitle

\section{Introduction}

Nonlinear models in (1+1)-dimensional space-time play important
hole both in field theory and in quantum mechanics. Some
of such models possess defect solutions representing domain walls that
appear, for example, in high energy physics \cite{1} and condensed
matter \cite{2}.

In this work we extend the method introduced in Ref.~{\cite{3}} to
present new potentials bearing topological (kinklike) or non
topological (lumplike) solutions. For example, we use the extended
deformation procedure to build a semi-vacuumless model, and the
corresponding domain wall which serves as seed for generation of
``dimension bubbles," as proposed in Refs.~{\cite{BG,mor}}. We
also show how to deform models having lumplike solutions, relevant
to the discussion of tachyonic excitations, to generate new ones
presenting topological solutions.

To begin, consider a theory of a single scalar real field in a
(1+1)-dimensional space-time, described by the Lagrangian density
\begin{equation}
{\cal L}=\frac{1}{2} \partial _\mu \phi \partial ^\mu \phi -
V(\phi ) \, . \label{lagran}
\end{equation}
We use the metric $(+,-)$, and we work with dimensionless fields
and coordinates, fields being defined in the whole space. The
equation of motion for static fields is
\begin{equation}
\frac{d^2\phi}{dx^2}=V^{\prime}(\phi ) , \label{eqm2}
\end{equation}
where the prime stands for the derivative with respect to the
argument. Mathematically, this corresponds to consider two-point
boundary value problems (with conditions imposed at $x_1=-\infty$
and $x_2=+\infty$) for second-order ordinary differential
equations \cite{PS}.

Consider the broad class of potentials having at least one
critical point $\bar\phi$ (that is, $V^{\prime}(\bar\phi)=0$), for
which $V(\bar\phi)=0$. In this case, solutions satisfying the
conditions
\begin{equation}
\lim_{x\rightarrow -\infty} \phi (x) = \bar \phi \,\,\,\,\,\, ,
\,\,\,\,\,\,\,
\lim_{x\rightarrow -\infty} \frac{d\phi}{dx} = 0 \, , \label{cond}
\end{equation}
obey the first order equation (a first integral of (\ref{eqm2}))
\begin{equation}
\frac{1}{2} \left( \frac{d\phi}{dx} \right)^2 = V(\phi(x)) \, ;
\label{eqfo}
\end{equation}
thus, such solutions equally share their energy densities between
gradient and potential parts.

Many important examples can be presented: the $\phi^4$-model, with
$V(\phi)=(1-\phi^2)^2/2$, is the prototype of theories having
topological solitons (kinklike solutions) connecting two minima,
$\phi(x)=\pm \tanh(x)$ in this case; a situation where non
topological (lumplike) solutions exist is the ``inverted
$\phi^4$-model", with potential given by $V(\phi)=\phi^2
(1-\phi^2)/2$, lumplike defects being $\phi(x) = \pm \, {\rm
sech}(x)$. One notice that the potential need not be nonnegative
for all values of $\phi$ but the solution must be such that
$V(\phi(x)) \geq 0$ for the whole range $-\infty < x < +\infty$.

\section{The deformation procedure}

Both topological and non topological solutions can be deformed,
according to the prescription introduced in \cite{3}, to generate
infinitely many new soluble problems. This method can be stated in
general form as the following theorem:

\begin{quote}
\noindent{\it Let $f=f(\phi)$ be a bijective function having
continuous non-vanishing derivative. For each potential $V(\phi)$,
bearing solutions satisfying conditions (\ref{cond}) (or
equivalently Eq. (\ref{eqfo})), the $f$-deformed model, defined by
\begin{equation}
{\wt V}(\phi) = \frac{V[f(\phi)]}{[f^{\prime}(\phi)]^2} \, ,
\label{Vtil}
\end{equation}
possesses solution given by
\begin{equation}
{\wt \phi}(x) = f^{-1}(\phi(x)) \, , \label{phitil}
\end{equation}
where $\phi(x)$ is a solution of the static equation of motion for
the original potential $V(\phi)$.}
\end{quote}

To prove this assertion, notice that the static equation of motion
of the new theory is written in terms of the old potential as
\begin{equation}
\frac{d^2\phi}{dx^2} =
\frac{1}{f^{\prime}(\phi)}V^{\prime}[f(\phi)] -
2V[f(\phi)]\frac{f^{\prime \prime}(\phi)}{[f^{\prime}(\phi)]^3} \,
. \label{dd1}
\end{equation}
On the other hand, taking the second derivative with respect to
$x$ of Eq. (\ref{phitil}), one finds
\begin{equation}
\frac{d^2{\wt \phi}}{dx^2} = \frac{1}{f^{\prime}({\wt \phi})}
\frac{d^2\phi}{dx^2} - \frac{f^{\prime \prime}({\wt
\phi})}{[f^{\prime}({\wt \phi})]^3} \left( \frac{d\phi}{dx}
\right)^2 \, . \label{dd2}
\end{equation}
It follows from (\ref{eqm2}), (\ref{eqfo}) and (\ref{phitil}) that
$d^2\phi/dx^2 = V^{\prime}[f({\wt \phi})]$ and $(d\phi / dx)^2 = 2
V[f({\wt \phi})]$ so that $\wt \phi$ satisfies (\ref{dd1}), as
stated. The energy density of the solution (\ref{phitil}) of the
$f$-deformed potential (\ref{Vtil}) is given by
\begin{equation}
{\wt \varepsilon}(x)=\left. \left( \frac{df^{-1}}{d\phi}
\right)^2 \left(\frac{d\phi}{dx}\right)^2 \right|_{\phi=
\phi_\pm (x)}\, .
\label{edens}
\end{equation}

Naturally, the deformation procedure heavily depends on the
deformation function $f(\phi )$. Assume that $f:{\bf R}\rightarrow
{\bf R}$ is bijective and has no critical points. In this case,
the $f$-deformation (and the deformation implemented by its
inverse $f^{-1}$) can be applied successively and one can define
equivalence classes of potentials related to each other by
repeated applications of the $f$- (or the $f^{-1}$-) deformation.
Each of such classes possesses an enumerable number of elements
which correspond to smooth deformations of a representative one,
all having the same topological characteristics. The generation
sequence of new theories is depicted in the diagram below.
$$
\xymatrix{
\cdots & \widehat{\widehat{V}} \ar[d]& &\widehat{V}
\ar[d] \ar[ll]_{f^{-1}} & & V
\ar[d] \ar[ll]_{f^{-1}} \ar[rr]^{f}& & \wt{V} \ar[d] \ar[rr]^{f} & &
\wt{\wt{V}} \ar[d] & \cdots\\
\cdots & \widehat{\!\widehat{\phi}}_d & &\widehat{\phi}_d
\ar[ll]^{f}& &\phi_d \ar[ll]^{f} \ar[rr]_{f^{-1}}& & \wt{\phi}_d
\ar[rr]_{f^{-1}}& & \wt{\!\wt{\phi}}_d & \cdots}
$$

As an example not considered before, take the $\phi^6$-model. This
model, for which the potential $V(\phi)=\phi^2(1-\phi^2)^2/2$ has
three degenerated minima at $0$ and $\pm 1$, is important since it
allows the discussion of first-order transitions. It possesses
kinklike solutions, $\phi^+_{\pm}(x)=\pm\sqrt{\left[1+ \tanh
(x)\right]/2}$, $\phi^-_{\pm}(x)=\pm\sqrt{\left[1- \tanh
(x)\right]/2}$, connecting the central vacuum with the lateral
ones. Take $f(\phi)=\sinh(\phi)$ as the deforming function. The
sinh-deformed $\phi^6$-potential is
\begin{equation}
{\wt V}(\phi) = \frac{1}{2} \tanh^2 (\phi) \left[ 1 -
\sinh^2(\phi) \right]^2 \label{defpfi6}
\end{equation}
and the sinh-deformed defects are
\begin{eqnarray}
\label{solphi6+} {\wt \phi}^+_{\pm}(x)&=&\pm{\rm
arcsinh}\left(\sqrt{\frac{1}{2}\left[ 1+\tanh (x)\right] }
\right),
\\
{\wt \phi}^-_{\pm}(x)&=&\pm{\rm arcsinh} \left(
\sqrt{\frac{1}{2}\left[ 1-\tanh (x)\right] }
\right).\label{solphi6-}
\end{eqnarray}
Notice that, since $f^{\prime}(\phi)> 1$ for the sinh-deformation,
the energy of the deformed solutions is diminished with respect to
the undeformed kinks. The reverse situation emerges if one takes
the inverse deformation implemented with $f^{-1}(\phi)={\rm
arcsinh}(\phi)$.

Interesting situations arise if one takes polynomial functions
implementing the deformations. Consider
\begin{equation}
p_{2 n + 1}(\phi)=\sum_{j=0}^{n} c_j \phi^{2 j+1}\, ,
\label{poly}
\end{equation}
with $c_j > 0$ for all $0 \leq j \leq n$. These are bijective
functions from ${\bf R}$ into ${\bf R}$ possessing positive
derivatives. Fixing $n=0$ corresponds to a trivial rescaling of
the field. For $n=1$, taking $c_0=c_1=1$, one has $f(\phi)=p_3
(\phi)=\phi + \phi^3$ with inverse given by $f^{-1}(\phi)=
(2/\sqrt{3})\sinh [ \, {\rm arcsinh} (3 \sqrt{3} \phi /2)/3 \, ]$.
Thus, the $p_3$-deformed $\phi^4$ model, for which the potential
has the form
\begin{equation}
{\wt V}(\phi) = \frac{1}{2} \left( \frac{1 - \phi^2 - 2 \phi^4
-\phi^6} {1 + 3\phi^2} \right)^2 \, ,
\label{p3phi4}
\end{equation}
supports topological solitons given by
\begin{equation}
{\wt \phi}_{\pm}(x)= \pm \frac{2}{\sqrt{3}} \, \sinh \left[
\frac{1}{3} \, {\rm arcsinh} \left( \frac{3 \sqrt{3}}{2} \tanh (x)
\right) \right]\, . \label{solp3}
\end{equation}
Naturally, the inverse deformation can be implemented leading to
another new soluble problem. But if one takes $n \geq 2$, the
inverse of $p_{2 n + 1}$ can not be in general expressed
analytically in terms of known functions. This leads to
circumstances where one knows analytically solutions of potentials
which can not be expressed in term of known functions and,
conversely, one has well-established potentials for which
solitonic solutions exist but are not expressible in terms of
known functions. For example, take $f(\phi)=p_5 (\phi)=\phi +
\phi^3 + \phi^5$. The $p_5$-deformed $\phi^4$ model has potential
given by
\begin{equation}
{\wt V}(\phi) = \frac{1}{2} \left( \frac{1 - \phi^2 - 2 \phi^4 - 3
\phi^6 - 2 \phi^8 - \phi^{10}} {1 + 3\phi^2 + 5\phi^4} \right)^2
\, ,
\label{p5phi4}
\end{equation}
but its solutions ${\wt \phi}_{\pm}(x) = \pm p_5^{-1}(\tanh x)$
are not known analytically. On the other hand,
\begin{equation}
{\widehat \phi}_{\pm}(x) = \pm \left[ \tanh (x) + \tanh^3 (x) +
\tanh^5 (x) \right] \label{solp5}
\end{equation}
are topological solutions of the potential ${\widehat V}(\phi) =
(1 - [p_5^{-1}(\phi)]^2)^2/(2[p_5^{-1\, \prime}(\phi)]^2)$ which
can not be written in terms of known functions.

The procedure can also be applied to potentials presenting non
topological, lumplike, solutions which are of direct interest to
tachyions in String Theory \cite{Z,MZ,JV}. Take, for example, the
Lorentzian lump
\begin{equation}
\phi_l(x)=\frac{1}{x^2 + 1}
\label{Llump}
\end{equation}
which solves Eq. (\ref{eqm2}) for the potential
\begin{equation}
V(\phi) = 2 (\phi^3 - \phi^4) \,
\label{Lpot}
\end{equation}
and satisfies conditions (\ref{cond}). Distinctly of the
topological solitons, this kind of solution is not stable. In
fact, the `secondary potential', that appears in the linearized
Schr\"odiger-like equation satisfied by the small perturbations
around $\phi_l(x)$ \cite{Jackiw} is given by
\begin{equation}
U(x) = V^{\prime \prime}(\phi_l (x))=12\frac{x^2 - 1}{(x^2 +
1)^2}\, .
\label{SP}
\end{equation}
This potential, a symmetric volcano-shaped potential, has zero
mode given by $\eta_0(x) \approx \phi_l^{\prime}(x) = - 2 x/(x^2 +
1)^2$, which does not correspond to the lowest energy state since
it has a node. Deforming the potential (\ref{Lpot}) with
$f(\phi)=\sinh \phi$ leads to the potential ${\wt V}(\phi) = 2
\tanh^2 (\phi) \left[ \sinh (\phi) - \sinh^2 (\phi) \right]$ which
possesses the lumplike solution ${\wt \phi}_l (x) = {\rm arcsinh
\left[ (x^2 + 1)^{-1} \right]}$.

\section{The extended deformation prescription}

The deformation prescription is powerful. The conditions under
which our theorem holds are maintained if we consider a function
for which the contra-domain is an interval of ${\bf R}$, that is,
if we take $f:{\bf R} \rightarrow {\rm I} \subset {\bf R}$. In
this case, however, the inverse transformation (engendered by
$f^{-1}:{\rm I} \rightarrow {\bf R}$) can only be applied for
models where the values of $\phi$ are restricted to ${\rm I}
\subset {\bf R}$.

We illustrate this possibility by asking for a deformation that
leads to a model of the form needed in Ref.~{\cite{mor}},
described by a ``semi-vacuumless'' potential, in contrast with the
vacuumless potential studied in Ref.~\cite{CV,Bazeia}. Consider
the new deformation function $f(\phi)=1-1/\sinh(e^\phi)$, acting
on the potential $V(\phi)=(1-\phi^2)^2/2$. The deformed potential
is
\begin{equation}
\label{dpm} {\wt V}(\phi)=\frac12e^{-2\phi}{\rm sech}^2(e^{\phi})
\left( 2 \sinh(e^{\phi}) - 1\right)^2 \, ,
\end{equation}
which is depicted in Fig.~1. The kinklike solution is
\begin{equation}
{\wt\phi}(x)=\ln\Biggl[{\rm
arcsinh}\left(\frac{1}{1-\tanh(x)}\right)\Biggr] \, .
\end{equation}
The deformed potential (\ref{dpm}) engenders the required profile:
it has a minimum at $\bar\phi=\ln[{\rm arcsinh}(1/2)]$ and another
one at $\phi\to\infty$. It is similar to the potential required in
Ref.~{\cite{mor}} for the existence of dimension bubbles. The
bubble can be generated from the above (deformed) model, after
removing the degeneracy between $\bar\phi$ and $\phi\to\infty$, in
a way similar to the standard situation, which is usually
implemented with the $\phi^4$ potential, the undeformed potential
that we have used to generate (\ref{dpm}). A key issue is that
such bubble is unstable against collapse, unless a mechanism to
balance the inward pressure due to the surface tension in the
bubble is found. In Ref.~{\cite{mor}}, the mechanism used to
stabilize the bubble requires another scalar field, in a way
similar to the case of non topological solitons previously
proposed in Ref.~{\cite{FGGK}}. This naturally leads to another
scenario, which involves at least two real scalar fields.

\begin{figure}[h]
\includegraphics[{height=6.0cm,width=9.6cm}]{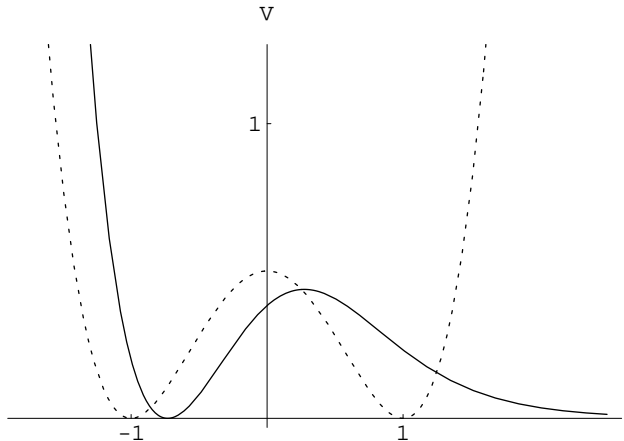}
\caption{The deformed potential ${\wt V}(\phi)$ of
Eq.~(\ref{dpm}), plotted as a function of the scalar field $\phi$;
the dashed line shows the potential of the undeformed $\phi^4$
model.}
\end{figure}

The deformation procedure can be extended even further, by
relaxing the requirement of $f$ being a bijective function, under
certain conditions. Suppose that $f$ is not bijective but it is
such that its inverse $f^{-1}$ (which exists in the context of
binary relations) is a multi-valued function with all branches
defined in the same interval $I\subset {\bf R}$. If the domain of
definition of $f^{-1}$ contains the interval where the values of
the solutions $\phi(x)$ of the original potential vary, then ${\wt
\phi}(x) = f^{-1}(\phi(x))$ are solutions of the new model
obtained by implementing the deformation with $f$. However, one
has to check out whether the deformed potential ${\wt
V}(\phi)=V[f(\phi)]/(f^{\prime}(\phi))^2$ is well defined on the
critical points of $f$. In fact, this does not happen in general
but occurs for some interesting cases.

Consider, for example, the function $f(\phi) = 2 \phi^2 - 1$; it
is defined for all values of $\phi$ and its inverse is the double
valued real function $f^{-1}(\phi)=\pm\sqrt{(1+\phi)/2}$, defined
in the interval $[-1,\infty)$. If we deform the $\phi^4$ model
with this function we end up with the potential ${\wt
V}(\phi)=\phi^2(1-\phi^2)^2/2$. The deformed kink solutions are
given by ${\wt\phi}(x)=\pm\sqrt{(1+\phi(x))/2}$ with $\phi(x)$
replaced by the solutions ($\pm\tanh(x)$) of the $\phi^4$ model,
which reproduce the known solutions of the $\phi^6$ theory. The
important aspect, in the present case, is that the $\tanh$-kink
corresponds to field values restricted to the interval $(-1,+1)$
which is contained within the domain of definition of the two
branches of $f^{-1}(\phi)$. The fact that the $\phi^6$ model can
be obtained from the $\phi^4$ potential in this way is
interesting, since these models have distinct characteristics.
Notice that the critical point of $f$ at $\phi=0$ does not disturb
the deformation in this case; this always occur for potentials
having a factor $(1-\phi^2)$, since the denominator of ${\wt
V}(\phi)$ (Eq.~\ref{Vtil}) is cancelled out. One can go on and
apply this deformation to the $\phi^6$ model; now, one finds the
deformed potential
\begin{equation}
\label{phi10}{\wt
V}(\phi)=\frac{1}{2}\phi^2(1-\phi^2)^2(1-2\phi^2)^2 \, ,
\end{equation}
with solutions given by
\begin{equation}
\label{sphi10}{\wt\phi}(x) =
\pm\sqrt{\frac{1}{2}\left(1\pm\sqrt{\frac{1}{2}\left[1\pm\tanh(x)\right]}
\right)} \, ,
\end{equation}
corresponding to kinks connecting neighbouring minima (located at
$-1$, $-1/\sqrt{2}$, $0$, $1/\sqrt{2}$ and $1$) of the potential
(\ref{phi10}). This potential is illustrated in Fig.~2. Repeating
the procedure for the potential (\ref{phi10}), the deformed
potential is a polynomial function of degree $18$, having sixteen
kink solutions connecting adjacent minima of the set $\left\{0,\,
\pm 1,\, \pm \sqrt{2}/2,\, \pm \sqrt{2+\sqrt{2}}/2,\, \pm
\sqrt{2-\sqrt{2}}/2\right\}$, and so on and so forth.

\begin{figure}[h]
\includegraphics[{height=6.0cm,width=9.6cm}]{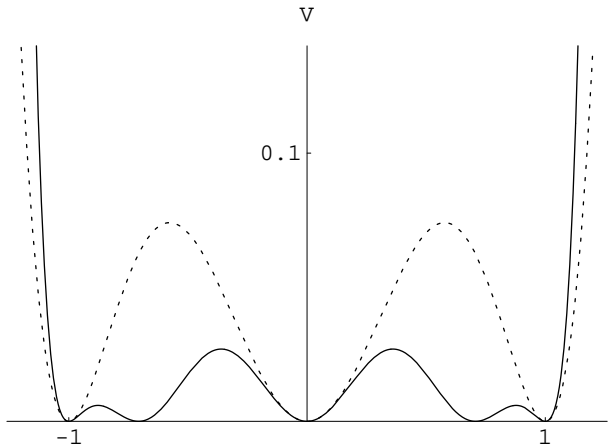}
\caption{The deformed potential ${\wt V}(\phi)$ of
Eq.~(\ref{phi10}), plotted as a function of the scalar field
$\phi$; the dashed line shows the potential of the undeformed
$\phi^6$ model.}
\end{figure}

The deformation implemented by the function $f(\phi)=2\phi^2 - 1$
can also be applied to a potential possessing lumplike solutions.
Consider the inverted $\phi^4$ potential
$V(\phi)=\phi^2(1-\phi^2)/2$, which has the lump solutions
$\phi(x)=\pm{\rm sech}(x)$. The deformed potential, in this case,
is given by
\begin{equation}
{\wt V}(\phi)=\frac{1}{8}(2\phi^2 - 1)^2(1 - \phi^2) \, .
\label{Vlump2}
\end{equation}
This potential, which is also unbounded from below, vanishes for
$\phi = \pm\sqrt{2}/2, \,\pm 1$, has an absolute maximum at
$\phi=0$ and local minima and maxima for $\pm\sqrt{2}/2$ and
$\pm\sqrt{5/6}$, respectively. Figure 3 shows a plot of this
potential. If we take $\phi(x)=+{\rm sech}(x)$ as the original
lump, then the deformed solutions obtained are given by
\begin{equation}
{\wt \phi}(x)=\pm\sqrt{\frac{1}{2}\left[1 + {\rm sech}(x)\right]}
\, , \label{Slump2}
\end{equation}
corresponding to lumps which start in the local minima (for
$x=-\infty$), go to the lateral zeros (when $x=0$) and come back
to the same minima (for $x=+\infty$) of the potential
(\ref{Vlump2}). On the other hand, if the undeformed lump is
$\phi(x)=-{\rm sech}(x)$, taking naively $f^{-1}(-{\rm sech}(x))$
leads to functions that do not have derivative at $x=0$ and,
therefore, are not acceptable solutions of the equations of
motion. In fact, the deformation of the $-{\rm sech}(x)$ lump
results in deformed kinks given by
\begin{equation}
{\wt \phi}(x)=\pm \left\{
\begin{array}{c}
-\sqrt{\frac{1}{2}\left[1 - {\rm sech}(x)\right]} \,\,\,\,\,\,\, x\leq 0 \\
+\sqrt{\frac{1}{2}\left[1 - {\rm sech}(x)\right]} \,\,\,\,\,\,\,
x\geq 0
\end{array}
\right. \,\, . \label{slumpkink}
\end{equation}
Each of such solutions (e.g., the $(+)$ one) starts at a minimum
($\phi = -\sqrt{2}/2$) when $x=-\infty$, goes to the absolute
maximum ($\phi=0$) at $x=0$ running along one of the branchs of
$f^{-1}$ (the lower branch) and, passing continuously to the other
branch (the upper one), reaches the other local minimum ($\phi =
+\sqrt{2}/2$) when $x=+\infty$. One notice that, again, the number
of solutions duplicates using such a deformation, whose inverse is
a double-valued function. But, in this case, novel topological
solutions emerge as deformations of a non topological one.

\begin{figure}[h]
\includegraphics[{height=6.0cm,width=9.6cm}]{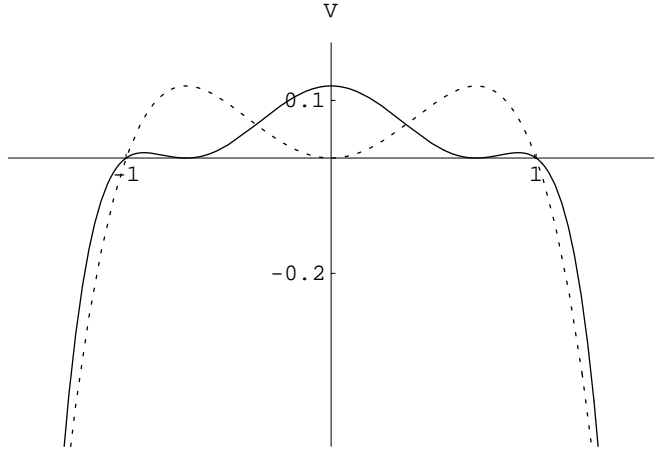}
\caption{The deformed potential ${\wt V}(\phi)$ of
Eq.~(\ref{Vlump2}), plotted as a function of the scalar field
$\phi$; the dashed line shows the potential of the undeformed
inverted $\phi^4$ model.}
\end{figure}

Potentials which have a factor $(1 - \phi^2)$ can also be deformed
using the function $f(\phi)=\sin(\phi)$, producing many
interesting situations. In fact, suppose the potential can be
written in the form
\begin{equation}
V(\phi)=(1 - \phi^2)\,U(\phi) \, ; \label{VU}
\end{equation}
this is always possible for all well-behaved potentials that
vanish at both values $\phi=\pm 1$, as shown by Taylor expansion.
Then, the sin-deformation leads to the potential
\begin{equation}
{\wt V}(\phi)=U[\sin(\phi)] \, , \label{VUD}
\end{equation}
which is a periodic potential, the critical points of $\sin(\phi)$
not causing any problem to the deformation process. The inverse of
the sine function is the infinitely valued function $f^{-1}(\phi)=
(-1)^k{\rm Arcsin}(\phi)+k\pi$, with $k\in {\bf Z}$ and ${\rm
Arcsin}(\phi)$ being the first determination of $\arcsin(\phi)$
(which varies from $-\pi/2$, for $\phi=-1$, to $+\pi/2$, when
$\phi=+1$), defined in the interval $\left[ -1,+1\right]$. So to
each solution of the original potential, whose field values range
in the interval $\left[ -1,+1\right]$, one finds infinitely many
solutions of the deformed, periodic, potential.

Consider firstly the $\phi^4$ model. Applying the sin-deformation
to it, one gets ${\wt V}(\phi)=\cos^2(\phi)/2$ which is one of the
forms of the sine-Gordon potential. The deformed solutions thus
obtained is given by ${\wt \phi}(x)=(-1)^k{\rm
Arcsin}\left[\pm\tanh(x)\right]+k\pi$, which correspond to all the
kink solutions (connecting neighbouring minima) of this
sine-Gordon model. For example, the kink solutions $\pm\tanh x$,
which connect the minima $\phi=\pm 1$ of the $\phi^4$ model in
both directions, are deformed into the kinks $\pm{\rm
Arcsin}\left[\tanh (x)\right]=2{\rm Arctan}(e^{\pm x}) - \pi /2$
(which runs between $-\pi /2$ and $\pi /2$) if one takes $k=0$
while, for $k=1$, the resulting solutions connect the minima $\pi
/2$ and $3\pi /2$ of the deformed potential.

This example can be readily extended to other polynomial
potentials, leading to a large class of sine-Gordon type of
potentials. For instance, the $\phi^6$ model,
$V(\phi)=\phi^2(1-\phi^2)^2/2$, deformed by the sine function,
becomes the potential
\begin{equation}
{\wt V}(\phi)=\frac{1}{2}\cos^2(\phi)(1 - \cos^2(\phi)) \, ,
\label{V6sin}
\end{equation}
having kink solutions given by
\begin{equation}
{\wt \phi}(x)=\pm (-1)^k{\rm Arcsin}\left(\sqrt{\frac{1}{2}\left[1
\pm \tanh (x)\right]}\right) + k\pi \, . \label{S6sin}
\end{equation}
If, on the other hand, one considers the potential $V(\phi)=(1 -
\phi^2)^3/2$, which is unbounded below and supports kinklike
solutions connecting the two inflection points at $\pm 1$, one
gets the potential
\begin{equation}
{\wt V}(\phi)=\frac{1}{2}\cos^4(\phi) \, , \label{V6I}
\end{equation}
with solutions given by
\begin{equation}
{\wt \phi}(x)=\pm (-1)^k{\rm Arcsin}\left( \frac{x}{\sqrt{1 +
x^2}} \right) + k\pi \, . \label{S6I}
\end{equation}

A particularly interesting situation appears if one consider the
inverted $\phi^4$ model, which presents lumplike solutions. The
sin-deformation of the potential $V(\phi)=\phi^2(1 - \phi^2)/2$
leads to the potential ${\wt V}(\phi)=\sin^2(\phi)/2$. In this
case, the lump solutions of $V(\phi)$, namely $\phi(x)=\pm {\rm
sech}(x)$, are deformed into ${\wt \phi}(x)=\pm (-1)^k{\rm
Arcsin}\left[{\rm sech}(x)\right]+k\pi$. Consider the
$(+)$-solution and take initially $k=0$. As $x$ varies from
$-\infty$ to $0$, ${\rm sech}(x)$ goes from $0$ to $1$, and ${\rm
Arcsin}\left[{\rm sech}(x)\right]=2{\rm Arctan}(e^x)$ changes from
$0$ to $\pi /2$. If one continuously makes $x$ goes from $0$ to
$+\infty$, then the deformed solution passes to the $k=1$ branch
of $\arcsin(\phi)$, $-{\rm Arcsin}\left[{\rm sech}(x)\right]+\pi$
($=2{\rm Arctan}(e^x)$ for $0\leq x < +\infty$), which varies from
$\pi /2$ to $\pi$ as $x$ goes from $0$ to $+\infty$. Thus, in this
case, the lump solution $+{\rm sech}(x)$ of the inverted $\phi^4$
model is deformed in the kink of the sine-Gordon model connecting
the minima $\phi=0$ and $\phi=\pi$. Under reversed conditions
(taking the $k=1$ branch before the $k=0$ one), the lump solution
$-{\rm sech}(x)$ leads to the anti-kink solution of the
sine-Gordon model running from the minimum $\phi=\pi$ to $0$. The
other topological solutions of the sine-Gordon model are obtained
considering the other adjacent branches of $\arcsin(\phi)$. This
is another remarkable example since one has non topological
solutions being deformed in topological ones.

Finally, we mention that many other soluble models can be
construct following the procedure presented in this paper. Also,
the discussion raised here might be extended to consider
situations where the fields are constrained to intervals of ${\bf
R}$, thus representing two-point boundary value problems defined
in finite or semi-infinite intervals of the $x$-axis. Such study
is left for future work.\\
\\

\noindent{\bf Acknowledgments}\\

The authors would like to thank W. Freire for comments, and CAPES,
CNPq, PROCAD and PRONEX for financial support. CAA thanks FUNCAP
for a fellowship.

\end{document}